\documentclass[showpacs,preprintnumbers,amsmath,amssymb]{revtex4}

\usepackage{graphicx}
\usepackage{dcolumn}
\usepackage{bm}

\begin{document}


\title{Model inspired by population genetics to study fragmentation of brittle plates}

\author{M. A. F. Gomes}
\email{mafg@ufpe.br}

\affiliation{%
Departamento de F\'{\i}sica, Universidade Federal de Pernambuco, 50670-901, Recife, PE, Brazil
}%

\author{Viviane M. de Oliveira}
\email{viviane@deinfo.ufrpe.br}

\affiliation{%
Departamento de Estat\'{\i}stica e Inform\'atica, Universidade Federal Rural de Pernambuco, 52171-900, Recife-PE, Brazil
}%

\date{\today}

\begin{abstract}
We use a model whose rules were inspired by population genetics, the random capability growth model, to describe the statistical details observed in experiments of fragmentation of brittle platelike objects, and in particular the existence of (i) composite scaling laws, (ii) small critical exponents $\tau$ associated with the power-law fragment-size distribution, and (iii) the typical pattern of cracks. The proposed computer simulations do not require numerical solutions of the Newton's equations of motion, nor several additional assumptions normally used in discrete element models. The model is also able to predict some physical aspects which could be tested in new experiments of fragmentation of brittle systems.
\end{abstract}


\maketitle

\section{INTRODUCTION}

The scientific and technologically important subject of failure, fracture, and fragmentation in condensed matter physics has a rich phenomenology and depends on a broad range of physical and chemical factors such as material composition, impurities, defects, internal and phase boundaries, and temperature, among others. Because of the complex interactions involving these factors, in most cases it has been difficult to develop a consistent understanding of fragmentation processes on a fundamental level. To complicate matters further, statistical complexities arise in experiments on fragmentation \cite{Grady}. For instance, the experiments are invariably statistically inhomogeneous; i. e. the fragment size varies as a function of position within the fragmented object because of the spatial variation of forces and energies causing fragmentation. The concept of scaling is important in a number of fragmentation processes \cite{Kaline}, and critical exponents can be precisely defined in several cases, depending \cite{Oddershede}, or not \cite{Meibon} on the shape of the object. More recently, the fragmentation of low-dimensional brittle materials embedded in a higher dimensional space has been studied by several groups \cite{Wittel, Katsuragi, Audoly, Linna, Katsuragi2, Kadono, Kun}.

In this work we use the random capability growth (RCG) model, a simple computer simulation whose rules are inspired by population genetics and which was recently introduced to study the evolution of the linguistic diversity on the Earth \cite{Oliveira, Oliveira2}, to describe the statistical details observed in experiments dealing with brittle platelike objects fragmented under several types of impact \cite{Meibon, Oddershede}. In impulsive fragmentation, the RCG model includes the effects of inhomogeneity and randomness of the brittle material, as well as the effect of the diffusion of the stress pulse causing the fragmentation. Furthermore, the RCG model predicts new physical aspects that could be tested in experiments. 

\section{THE MODEL}

The RCG model is defined on a two-dimensional lattice composed of $A=L^2$ sites with periodic boundary conditions, which simulates a two-dimensional plate whose height $h << L$. We assume that due to the impact there is a stress pulse that diffuses or disperses in the system starting from the impact point \cite{Withers}. During this diffusion process some connections between lattice sites are broken depending on their particular mechanical strength. As we are assuming a disordered non-crystalline medium, this mechanical strength is a randomly distributed variable on the lattice. Thus, to each lattice site $i$ we ascribe a given capability $C_i$ whose value we estimate from a uniform distribution in the interval 0-1, and so $C_i$ refers to the facility to accomodate at site $i$ an excess of stress from the original impact.

\begin{figure*}
\includegraphics[scale=0.22]{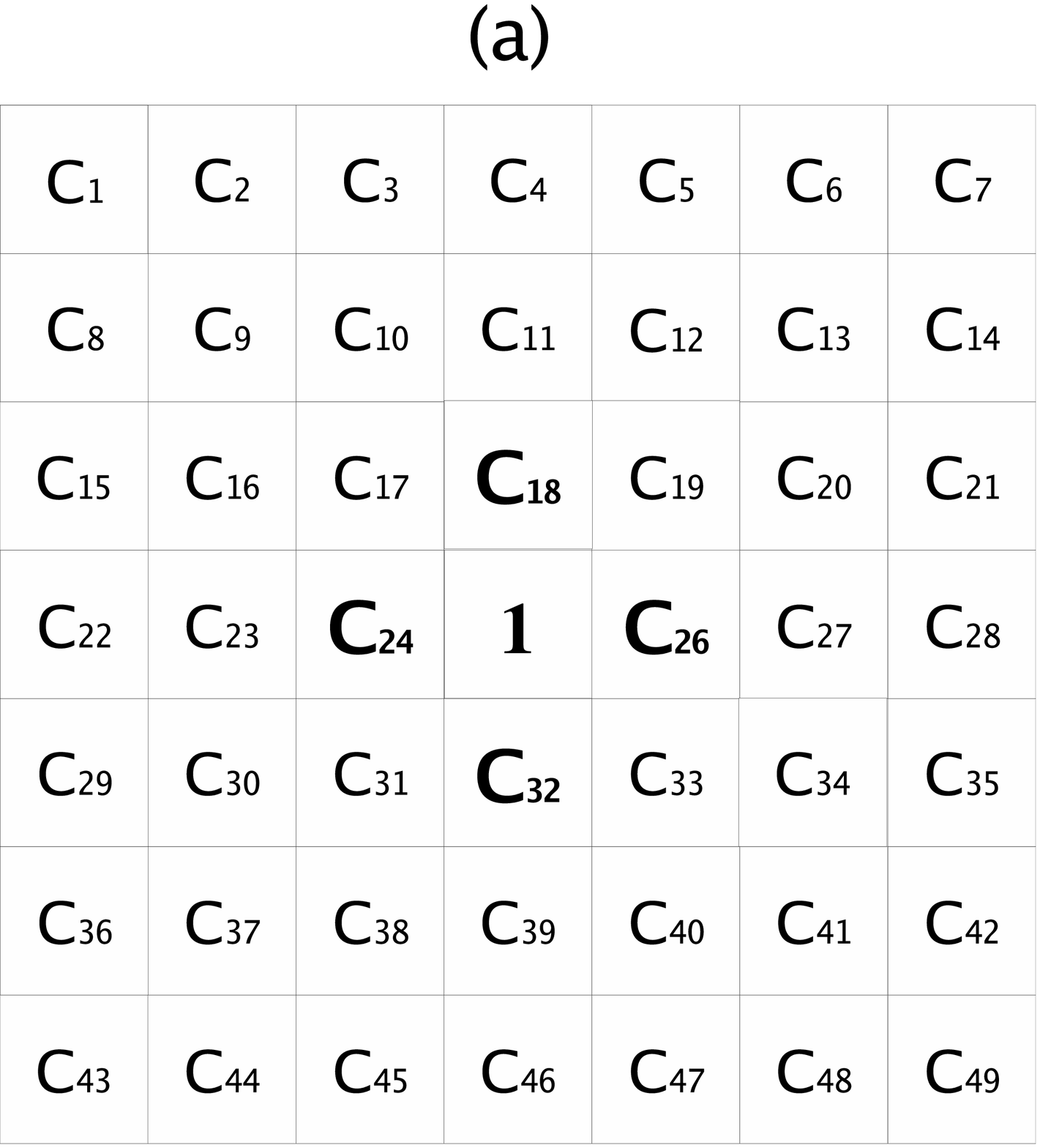}
\includegraphics[scale=0.22]{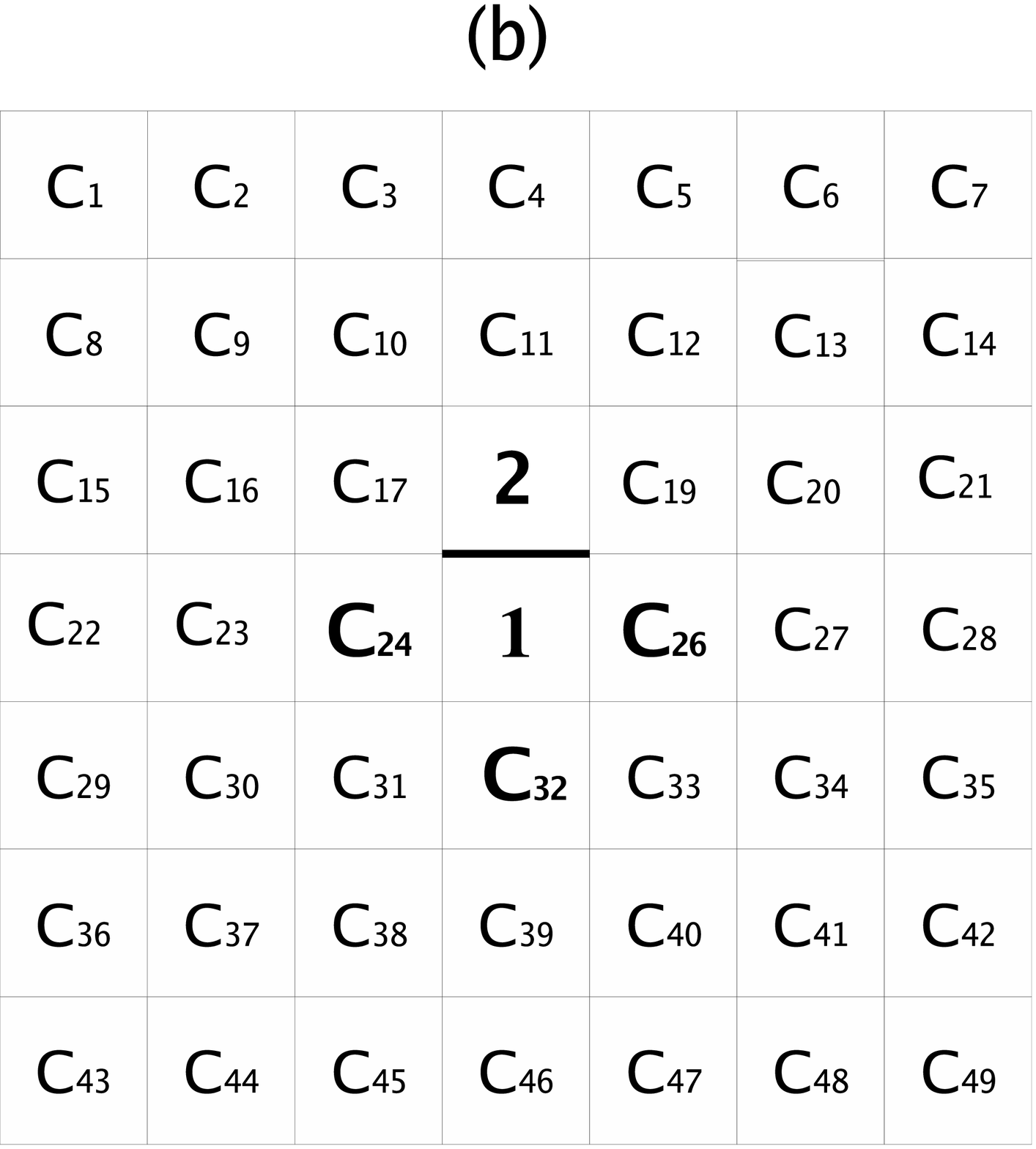}
\includegraphics[scale=0.22]{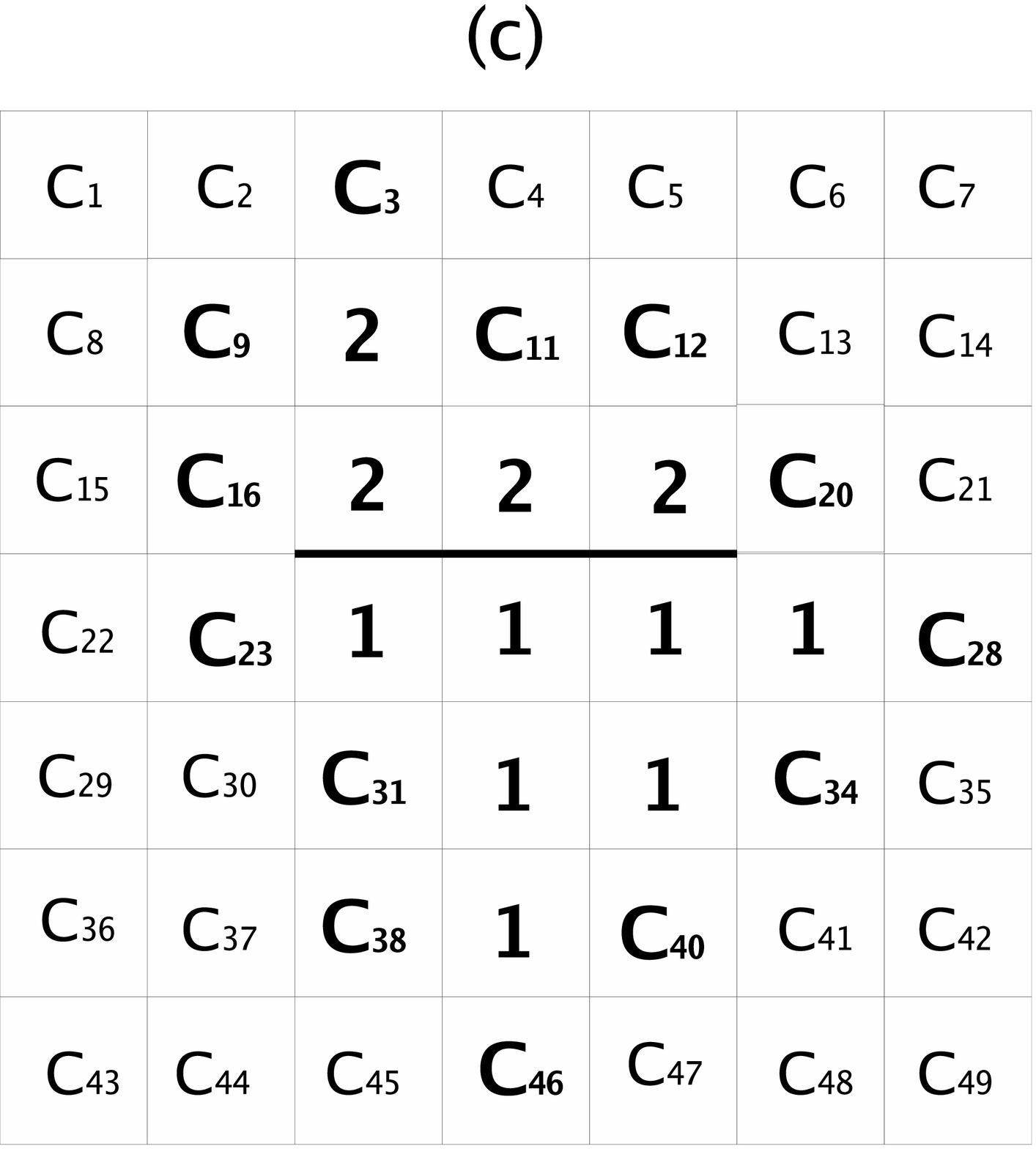}
\includegraphics[scale=0.22]{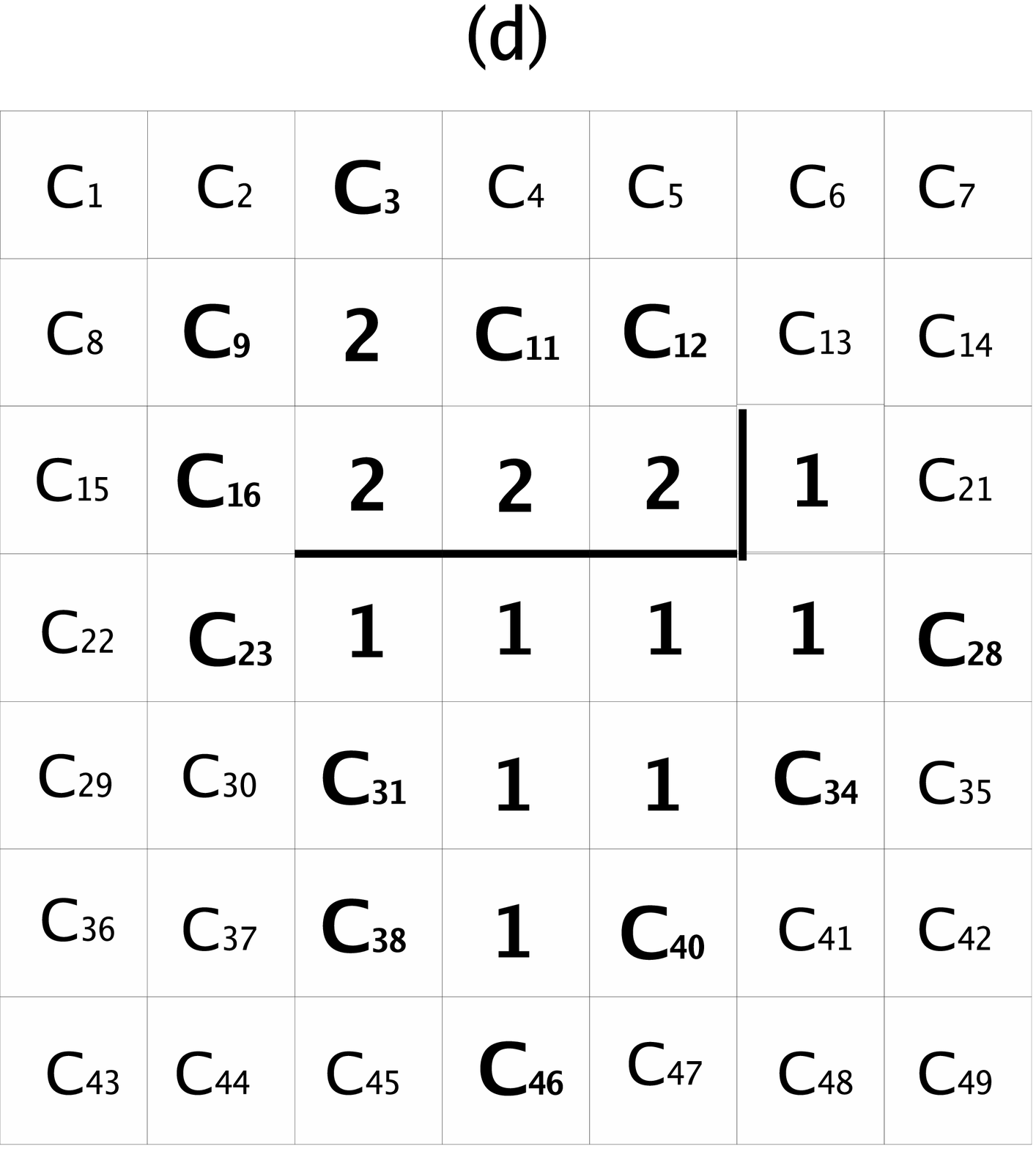}
\includegraphics[scale=0.22]{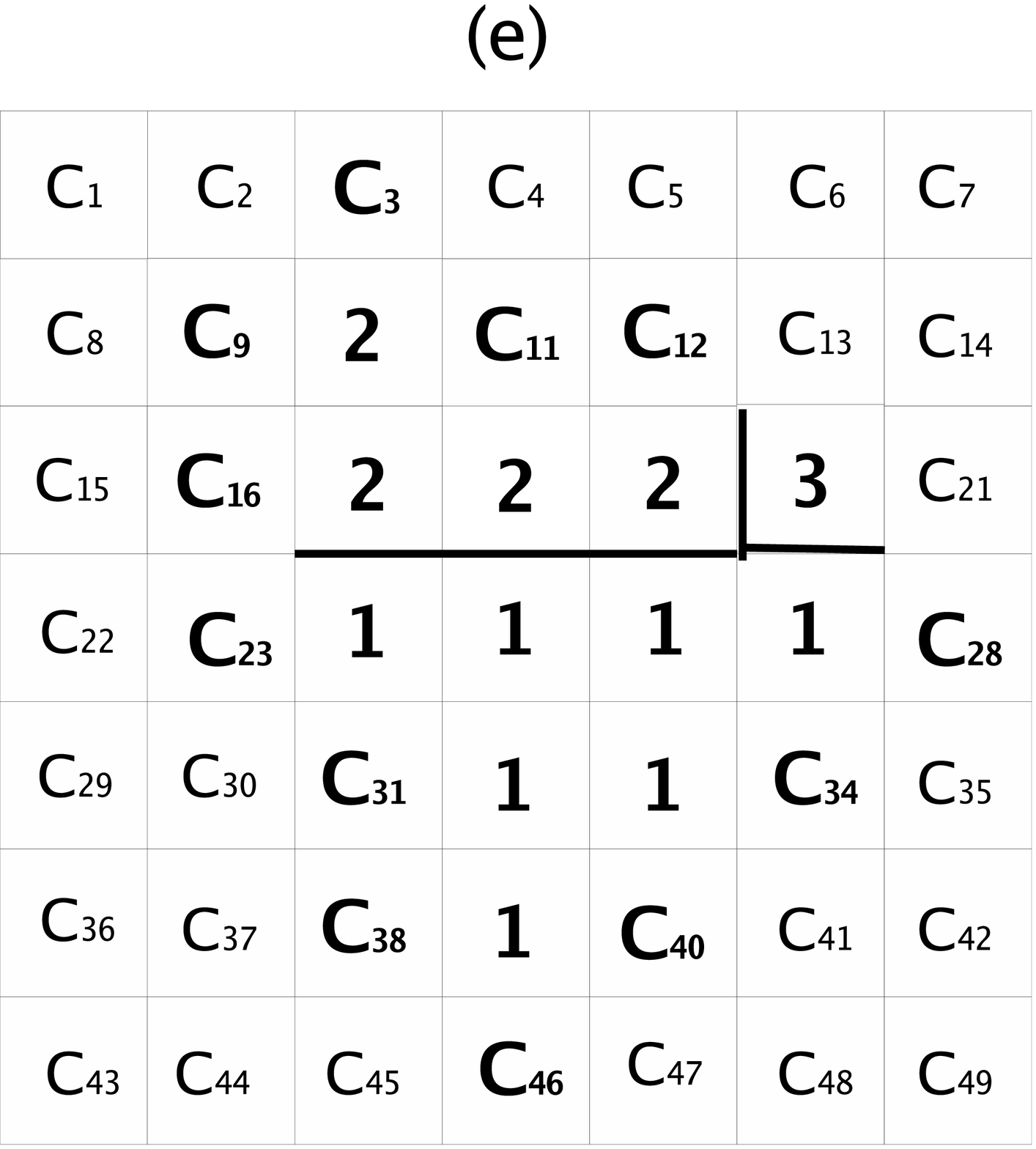}
\caption{In a lattice composed by $A=7^2$ sites we show: (a) Impact site (labeled by number 1) and its four nearest neighbors (time $t=1$); (b) The occurrence of a mutation begining the nucleation of a new(future) fragment labeled by the integer 2 (time $t=2$); (c) Cluster whose sites were visited by the stress pulse and its boundary at time $t=11$; (d) Increasing of the fragment labeled by the integer 1 (time $t=12$); (e) The begining of the nucleation of a new fragment which is labeled by the integer 3 (time $t=12$).}
\end{figure*}

\begin{figure}
\vspace{-0.5cm}
\centering
\includegraphics[height=8cm,width=9cm,angle=0]{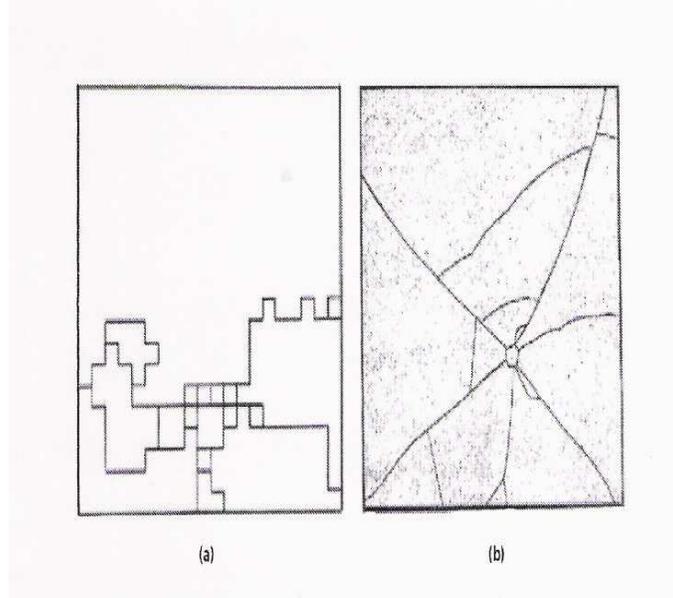}
\caption{(a) Snapshot of a particular realization of the dynamics after all sites be visited by the stress pulse. The lattice size is $L=20$ and $\alpha=0.6$. (b) Typical distribution of fragments observed in a glazed-tile after impact with a conic steel projectile falling in free fall. See text for detail.}
\end{figure}

\begin{figure}
\vspace{-0.5cm}
\centering
\includegraphics[height=9cm,width=9cm,angle=270]{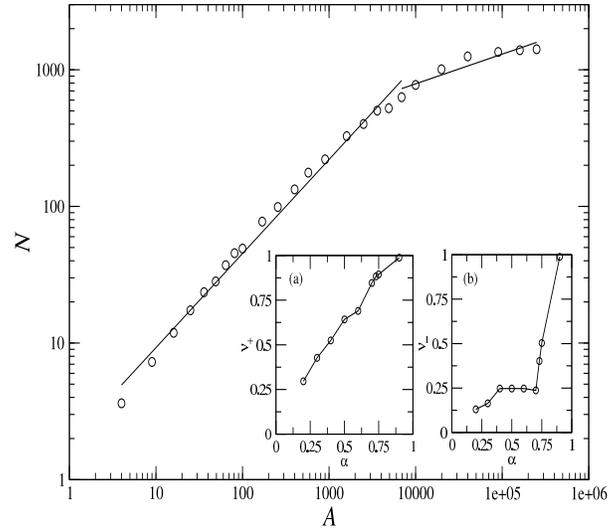}
\vspace{-0.8cm}
\caption{ Number of fragments $N$ as a function of the area $A$ for $\alpha=0.6$. The exponents are $\nu=\nu_{+}=0.69 \pm 0.01$ for $4<A<7000$, and $\nu=\nu_{-}=0.22 \pm 0.01$ for $7000<A<250,000$. In the inset we present exponent $\nu$ as a function of $\alpha$ for (a) small and intermediate areas and (b) large areas.}
\end{figure}

\begin{figure}
\vspace{-0.59cm}
\centering
\includegraphics[height=9cm,width=9cm,angle=270]{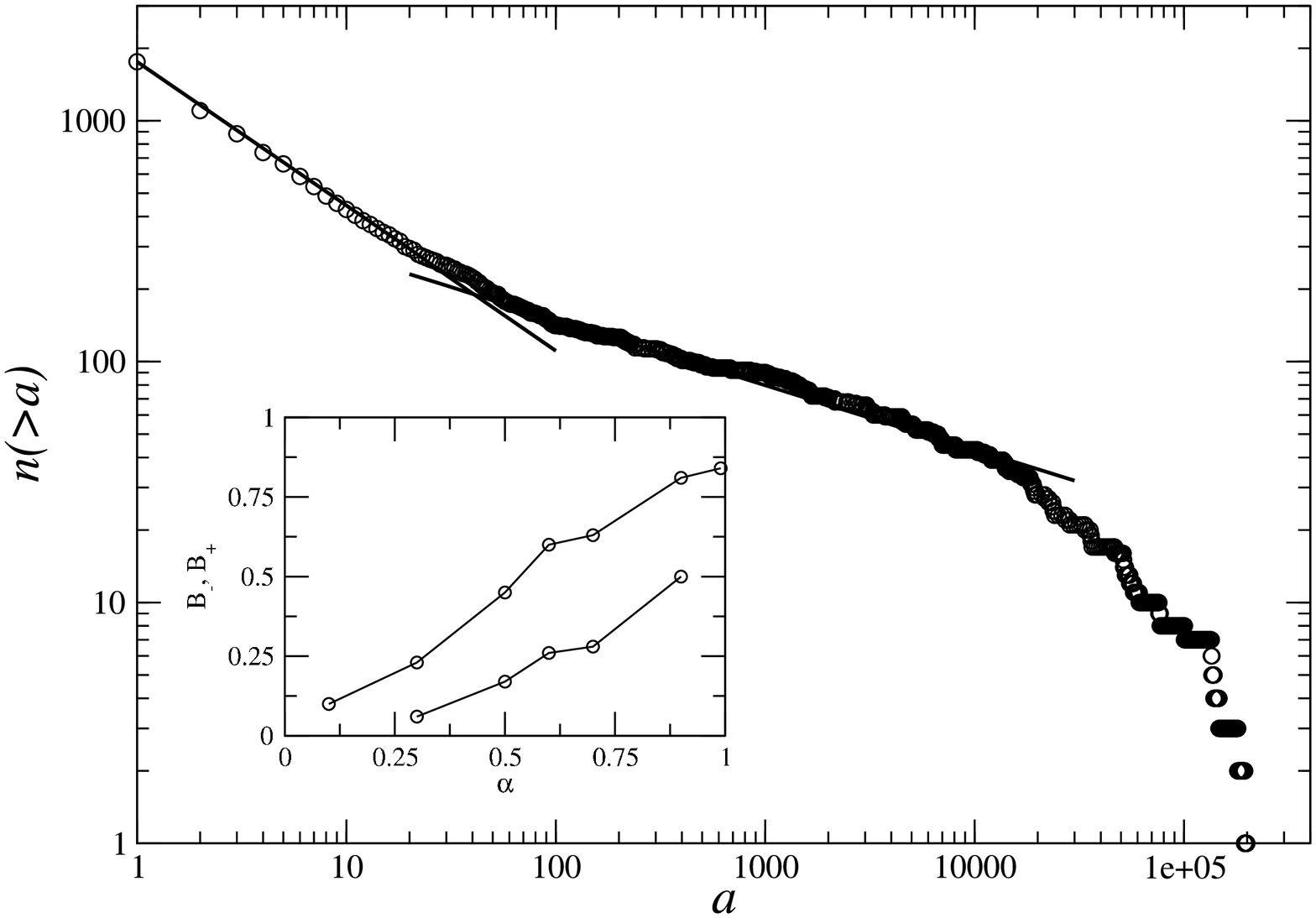}
\vspace{-0.8cm}
\caption{Number ~ of fragments with area greater than $a$, $n(>a)$, ~ as a function of $a$, for $\alpha=0.6$, and $L=500$. $n(>a) \sim a^{-B}$ with $B=B_{+}=0.60 \pm 0.01$ for $1<a<20$ and $B=B_{-}=0.27 \pm 0.01$ for $45<a<12,000$. In the inset we present the exponents $B_{-}$ (bottom) and $B_{+}$ (top) as a function of $\alpha$.}
\end{figure}

In the first step of the dynamics, we randomly choose one site where the excess of stress due to the impact is firstly injected. This site will belong to the fragment labeled by number 1. In the second step, one of the four nearest neighbors of this site will be chosen to receive the stress pulse with probability proportional to $C_i$ (see Figure 1a). We consider that there is a finite probability $p$ to occur a {\it mutation} which will break a single connection between two sites and which will contribute to the formation of a new fragment at a future time. In the context of population genetics mutations are the mechanisms responsible for generating diversity. In the present case, if a mutation occurs the chosen neighbor site is labeled by number 2 and the nucleation of a new fragment begins (Figure 1b). Otherwise, it is labeled by number 1 and the size of the initial fragment is increased. The mutation probability is given by $p=\alpha / \Sigma$, where $\alpha$ is a constant in the interval 0-1, and $\Sigma$ is the {\it fitness} of the fragment, defined as the sum of all the capabilities of the sites belonging to that specific fragment. Therefore, the initial fitness of the first (impact) site is the capability of the initial site. This rule for the mutation probability was inspired by population genetics where the most adapted organisms are less likely to mutate than poorly adapted organisms \cite{Barton}. The increase of the parameter $\alpha$ introduces an increment in the mutation probability, that is, $\alpha$ is a direct measure of the impact energy.

In the subsequent steps, we check the sites which are on the boundary of the cluster whose sites were visited by the stress pulse, and we choose one of those sites according to their capabilities (Figure 1c). Those with higher capabilities have a higher likelihood to be visited by the stress pulse. We then choose the fragment to which this site will belong among their neighboring fragments. The fragments with higher fitness have a higher chance to increase in size (Figure 1d). As before, the mutation probability is inversely proportional to the fitness of the chose fragment and if a mutation occurs the site is labeled by number 3 (Figure 1e). This process will continue until all sites be visited by the pulse. At this point, we verify the total number of fragments $N$. Each selection of a site corresponds to an increment of one time unit.

With the passage of the time, the probability increases to find fragments with increasing fitness, thus when the distance from the (initial) site of impact grows, the size of the fragments increases as well, as physically expected. Figure 2a shows a particular configuration of fragmentation on a lattice with $20^2$ sites. For comparison, Figure 2b shows a typical distribution of fragments observed in a glazed-tile of size $150 \times 150 mm^2$ and thickness $5 mm$, after impact with a conic steel projectile of $238g$ of mass falling in free fall from a height of $5 m$ \cite{Valdemiro}. 

\section{RESULTS AND DISCUSSION}

Due to the effect of increasing fitness for large times, discussed in the previous section, if we fix the impact energy ($\alpha$ fixed), and for plates of increasing areas, we observe that the number of fragments per area, $N/A$, is a decreasing function of $A$. This is exemplified in Figure 3 which shows a composite power-law for $N(A)$ for a simulation with $\alpha=0.60$, and average on 200 similar experiments for $L \le 300$, and on 20 similar experiments for $L=400$, and 500:

\begin{figure}
\vspace{-0.8cm}
\centering
\includegraphics[height=8cm,width=8cm,angle=270]{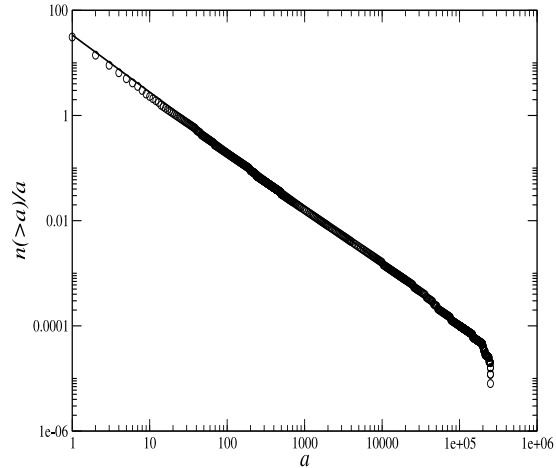}
\vspace{-0.8cm}
\caption{ $n(>a)/a$ as a function of $a$ for $\alpha=0.1$ and $L=500$. $n(>a)/a \sim a^{-\tau}$ with $\tau=1.08 \pm 0.01$ for $1<a<30,000$.}
\end{figure}

\begin{figure}
\centering
\includegraphics[height=8.5cm,width=8cm,angle=270]{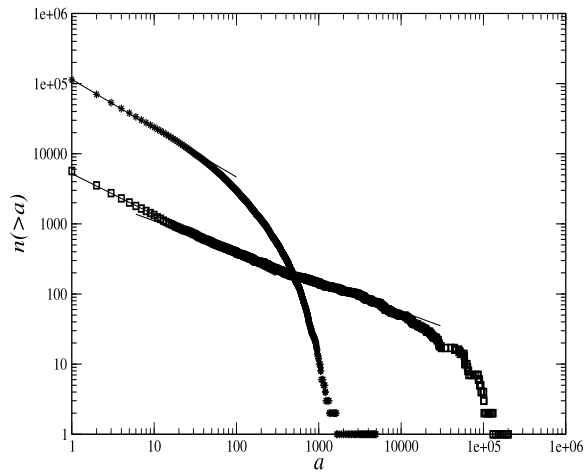}
\vspace{-0.8cm}
\caption{Number of fragments with area greater than $a$, $n(>a)$ as a function of $a$ for $\alpha=0.6$ and $L=500$ for $f_{max}=1000$ $(\square)$ and $f_{max}=10$ $(\star)$. When $f_{max}=1000$ we estimate $B_{+}=0.61 \pm 0.01$ for $1<a<30$ and $B_{-}=0.45 \pm 0.01$ for $30<a<10,000$. When $f_{max}=10$, we have $B_{+}=0.71 \pm 0.01$ for $1<a<30$.}
\end{figure}

\begin{figure}
\vspace{-0.5cm}
\centering
\includegraphics[height=8cm,width=8cm,angle=270]{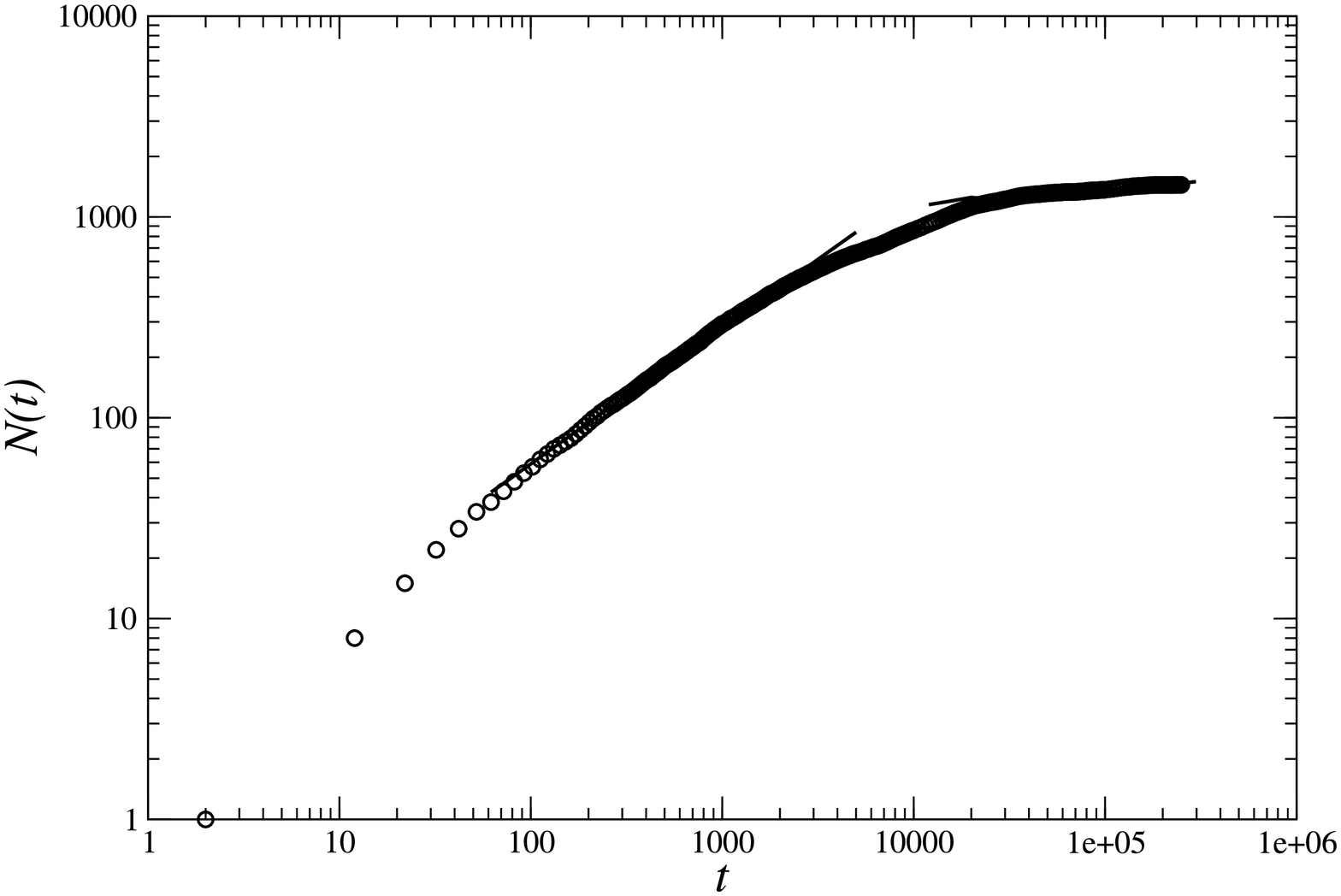}
\vspace{-0.8cm}
\caption{ Evolution of $N(t)$ for $\alpha=0.6$ and $L=500$. In this curve we have $N(t) \sim t^{0.68\pm 0.01}$ for $60<t<2,100$ and $N(t) \sim t^{0.08\pm 0.01}$ for $24,000<t<250,000$.}
\end{figure}

\begin{equation}
N \sim A^{\nu},
\end{equation}

\noindent
with the exponent $\nu$ assuming the value $\nu_{+}=0.69 \pm 0.01$ for small and intermediate areas, and $\nu_{-}=0.22 \pm 0.01$ for large areas. The dependence of the exponent $\nu$ with the parameter $\alpha$ is shown in the inset of Figure 3. As $\alpha$ increases, $\nu$ increases towards the maximum value $\nu=1$ in an approximately linear way, and the composite scaling reduces to a single scaling when $\nu=1$. Unfortunately fragmentation experiments in general do not examine the dependence of the total number of fragments with the area of the plates. Thus after the experimental measurement of $N(A)$, we could compare the data with our prediction in equation 1, and Figure 3.

If we work with a scaling distribution of fragments of the type $n(a)=n(1) a^{-\tau}$ (where $n(1)$ is the number of fragments of minimun (unit) area), or an accumulated distribution of the type $n(>a)\sim a^{-B}$ ($B=\tau -1$), both defined in the interval $1\le a \le a_{max}$ on a 2D system of area $A=L^2$, we have the sum rule $A=\int n(a)\cdot a \cdot da = \left[ \frac{n(1)}{(2-\tau)} \right] (a_{max}^{2-\tau}-1)$. The last result reduces to $A \approx [n(1)/(2-\tau)]a_{max}^{2-\tau}$, (for $\tau<2$, and $a_{max} >>1$, as commonly observed in 2D real fragmentation processes for not too large impact energy). Furthermore, based on many experimental results it is valid to assume that the largest fragment size $a_{max}$ is proportional to the total area $A$ of the system (also for not too large impact energy); in this case we get from the last expression $n(1) \sim A^{\tau-1}$. Thus, we find that $N=\int n(s) ds \approx [n(1)/(\tau -1)] \sim A^{\tau -1} = A^B$, provided $\tau>1$; i.e. if the scaling (1) is valid, the expected differential distribution of fragments $n(a)$ tends to scale as $a^{-1-\nu}$ or, equivalently $n(>a) \sim a^{-\nu}$, with $\nu=B$.

The most frequently studied statistical function in fragmentation experiments is the average number of fragments of size larger than $s$, $n(>s)$, or area larger than $a$, $n(>a)$, for breakup of plates. The last statistical function for our fragmentation model is shown in Figure 4 for $\alpha=0.60$, and $L=500$ (10 similar experiments). In our model, $n(>a)$ exhibits in general a composite scaling law characterized by two different critical exponents $B_{+}$, and $B_{-}$ valid respectively for small and intermediate area fragments. Thus, the expected differential distribution of fragments, $n(a)$, which is the derivative of $n(>a)$ with respect to $a$, scales as $n(a) \sim a^{-\tau}$, with $\tau$ assuming the values $\tau_{+}=B_{+}+1$, and $\tau_{-}=B_{-}+1$, for small and intermediate areas, respectively. With the power law fit in Figure 4 we find $B_{+}=0.60 \pm 0.01$ ($\tau_{+}=1.60 \pm 0.01$), and $B_{-}=0.27 \pm 0.01$ ($\tau_{-}=1.27 \pm 0.01$). It can be noticed that the exponents $B$ (or $\tau$) obtained from Figure 4 are in agreement with the relation $B=\nu$ and with Figure 3. For the reader to develop a better insight into our model, we show $B$ as a function of $\alpha$ in the inset of Figure 4. We notice from this inset that both $B_{-}$ and $B_{+}$ present a nearly linear dependence with $\alpha$, with a constant difference ($B_{+}-B_{-}$).

The behavior observed in Figure 4 for $n(>a)$, with composite scaling laws, is in very good agreement with the experiments of Meibom and Balslev \cite{Meibon} of fragmentation of plates of dry clay falling onto a hard floor from a height of $2.0m$ (In \cite{Meibon} the experimental cumulative functions corresponding to $n(>a)$ are shown, although the critical exponents reported there, $\beta$, are those of the differential distribution, i.e. the same as $\tau$, in the standard notation adopted in the present work.). In the experiments of Meibom and Balslev, the observed values for the exponents were $\tau_{+}=1.57\pm 0.09$, and $\tau_{-}=1.19 \pm 0.08$, irrespective the aspect ratio of the plates, $h/L$, in the interval 0.015 to 0.133. It is interesting to observe that the exponent $\tau_{-}=1.27 \pm 0.01$ found in Figure 4 is only $5\%$ off from the exponent $\tau=1.35 \pm 0.02$ obtained in recent experiments of impact fragmentation of egg shells \cite{Wittel}. Furthermore, for $\alpha=0.10$, the accumulated distribution is shown in Figure 5. In this case, we observe the power law $n(>a)/a \sim a^{-B-1}=a^{-\tau}$, with $\tau=1.08\pm 0.01$ exactly as observed by Oddershede et al \cite{Oddershede} with the fragmentation of gypsum disks of $320mm$ in diameter and $5mm$ thick. In our Figure 5 the scaling persists over four decades in $a$, although in Figure 2 of \cite{Oddershede} the scaling occurs only over two decades due to the difficulty to detect experimentally those fragments with mass in the range $5 \times 10^{-5}g$ to $5\times 10^{-3}g$.

In order to understand the role of the brittleness of the material on the critical exponents, we assumed that the fitness of each fragment is bounded by a given maximum value, $f_{max}$, which is randomly chosen from an uniform distribution. In this way, the smaller $f_{max}$, the more brittle is the material. In Figure 6, we present the average number of fragments of area larger than $a$, $n(>a)$, for $\alpha=0.6$ and $L=500$, when $f_{max}=1000$ and $f_{max}=10$. When we compare these results with those shown in Figure 4 (where there is no limit to $f_{max}$), we notice that the exponent $B_{+}$ is roughly the same when $f_{max}=1000$, and a slight deviation of $10\%$ is obtained for $f_{max}$=10. This seems to indicate that $B_{+}$ is insensitive to the brittleness of the material. On the other hand, we see that $B_{-}$ varies with the brittleness. For instance, when $f_{max}=1000$ we estimate $B_{-}=0.45 \pm 0.01$, which represents a deviation of $50\%$ in comparison to the non-bounded case (or a deviation of $14\%$ in $\tau_{-} = 1+ B_{-}$), and for very small values of $f_{max}$, the scaling region described by $B_{-}$ does not exist.

Finally, Figure 7 shows the time evolution of the number of fragments, $N(t)$, for the same values of $\alpha$ and $L$ used in Figure 4 (average of 10 experiments). The time scaling appearing in Figure 7 could be tested in new experiments of fragmentation of brittle platelike objects using present-day high-speed high-resolution digital imaging techniques.

\section{SUMMARY AND CONCLUSION}

It can be observed that the RCG model is useful in obtaining not only the composite scaling laws observed in the fragmentation of brittle materials \cite{Meibon}, but also the small exponents $\tau$ (close to 1) \cite{Oddershede, Meibon, Wittel}, and the typical pattern of cracks observed experimentally (Figure 2). These achievements of the RCG model are obtained without resorting to out-of-plane bending modes and discrete element models where the dynamics of the breakup process are determined by molecular dynamics simulations, as in references \cite{Wittel} and \cite{Linna}. Thus it is pertinent to ask: Why indeed should the simple RCG model exhibit all these aspects observed in experiments of fragmentation with brittle materials? This is not an easy question. A possible answer, though, is suggested just by the two ingredients of the model: It could be said that (i) diffusion (in a disordered medium described by random capabilities) of a stress pulse due to the impact, and (ii) breakup probability stochastically controlled by the fitness function (as described in the fourth paragraph) are very important physical ingredients to be considered in the understanding of the basic statistical aspects of low-dimensional brittle fracture. 
It is interesting to note that of these two ingredients inspired by population genetics simulations, the ingredient (i) is physically plausible in a number of fragmentation processes, while the ingredient (ii) is more subtle, and deserves a further detailed investigation which is outside the scope of the present paper. 
Moreover, it is opportune to stress that the understanding of the physical origin of nonequilibrium structures like the crack patterns obtained in impulsive fragmentation remains one of the most challenging problems in statistical mechanics.

In conclusion, we have used a simple one-parameter fragmentation model which describes quite well the rich phenomenology observed in the fragmentation of brittle platelike systems, and in particular the existence of composite scaling laws. The model predicts new scaling laws, and we urge that more experiments be performed to investigate the laws proposed here.

We thank I. R. Tsang by the Figure 2a, V. P. Brito for provide us with his experimental data showed in Figure 2b and M. L. Sundheimer for proofreading the manuscript. V.~M. de Oliveira and M.~A.~F. Gomes are supported by Conselho Nacional de Desenvolvimento Cient\'{\i}fico e Tecnol\'ogico and Programa de N\'ucleos de Excel\^encia (Brazilian Agencies).

\end{document}